\documentclass[aps,pre,twocolumn,showpacs,superscriptaddress,noeprint]{revtex4-1}
\usepackage{epsfig}
\usepackage{times}
\usepackage{amsmath}
\usepackage{natbib}
\usepackage{graphicx}
\usepackage{color}
\usepackage[breaklinks=true]{hyperref}
\usepackage{breakcites}
\usepackage{hypernat}
\usepackage{float}
\usepackage{epstopdf}
\usepackage[normalem]{ulem}

\begin{document}

\title{Strategy equilibrium in dilemma games with off-diagonal payoff perturbations}

\author{Marco A. Amaral}
\email{marcoantonio.amaral@gmail.com}
\affiliation{Universidade Federal do Sul da Bahia - BA, Brazil}

\author{Marco A. Javarone}
\email{marcojavarone@gmail.com}
\affiliation{Department of Mathematics, University College London, London, UK}

\begin{abstract}
We analyse the strategy equilibrium of dilemma games considering a payoff matrix affected by small and random perturbations on the off-diagonal. 
Notably, a recent work~\cite{amaral_javarone_2020} reported that, while cooperation is sustained by perturbations acting on the main diagonal, a less clear scenario emerges when perturbations act on the off-diagonal. Thus, the second case represents the core of this investigation, aimed at completing the description of the effects that payoff perturbations have on the dynamics of evolutionary games.
Our results, achieved by analysing the proposed model under a variety of configurations, as different update rules, suggest that off-diagonal perturbations actually constitute a non-trivial form of noise. 
In particular, the most interesting effects are detected near the phase transition, as perturbations tend to move the strategy distribution towards non-ordered states of equilibrium, supporting cooperation when defection is pervading the population, and supporting defection in the opposite case. 
To conclude, we identified a form of noise that, under controlled conditions, could be used to enhance cooperation, and greatly delay its extinction.

\end{abstract}

\maketitle

\section{Introduction}
\label{Introduction}
Perturbative methods find large utilisation for studying a number of problems in physics, spanning from classical to quantum mechanics~\cite{broer01,bransden01}. 
Furthermore, as we know from the theory of chaos~\cite{smith01}, even small perturbations can have drastic effects on the dynamics of some systems, as those particularly sensible to initial conditions. The logistic map, for instance, is very useful for studying these phenomena. Also, the theory of chaos has the merit to have widely popularised fascinating results, as the so called 'butterfly effect', being nowadays mentioned both in books and movies (e.g.~\cite{movie01}), although sometimes without proper scientific care.

Perturbations have also been used in evolutionary game theory for studying the emergence of cooperation~\cite{Capraro2018,castellano_rmp09}, as recently reported in~\cite{amaral_javarone_2020}. 
Notably, the latter shows that small and random perturbations on the payoff matrix, of social dilemmas, can strongly influence the strategy equilibrium of a population. However, while their effect on the main diagonal has been clarified, i.e. they support cooperative behaviours, the effect given by off-diagonal perturbations is still unclear.
Thus, here we aim to clarify this specific aspect of the perturbative method presented in~\cite{amaral_javarone_2020}, and to obtain a complete description of the dynamics of evolutionary games, affected by small and random perturbations on their payoff matrix.
To this end, our analyses consider a variety of conditions, as three different update rules, and various games (e.g. Stag-Hunt, Snow-Drift, and Harmony Game). 
For the sake of clarity, and for making the manuscript self contained, the main model is fully described in the following section, with all relevant information.
Before moving further, let us give a brief introduction on the framework of Evolutionary Game Theory (hereinafter EGT) that we use for our investigation.

EGT is a suitable framework to study the emergence of cooperation among selfish individuals~\cite{Smith82, Nowak2006, Perc2017}. While cooperation still represents a lively challenge~\cite{Pennisi2005}, EGT has given many important insights on how such a phenomenon can spontaneously emerge at many scales in biological and social systems. Among the most studied mechanisms are kin selection~\cite{Hamilton1964}, direct and indirect reciprocity \cite{trivers_qrb71, axelrod_s81}, network reciprocity~\cite{Nowak1992a, wardil_epl09, wardil_pre10}, group selection~\cite{wilson_ds_an77} and heterogeneity~\cite{perc_bs10, Amaral2016, Amaral2015}.
Specifically, heterogeneity (sometimes dubbed as diversity) was found to be a fundamental element for supporting cooperative behaviours in evolutionary systems~\cite{santos_jtb12, wang_z_pre14b}. A group composed of individuals with different skills can easily become more effective in several challenges. At the same time, group diversity can shield the possible flaws of individual members while sustaining the whole group in a synergetic manner. This rationale has been applied to economies, enterprise communities, vaccination models, and even in biological evolution~\cite{Stewart2016, santos_jtb12, Qin2017a, zhang_xd_cpb12, RajibArefin2019}. Besides, that represents a fundamental aspect in genetic algorithms as well~\cite{Holland1992}, whose core mechanism is based on an heterogeneous population of candidate solutions that evolves towards a (sub-)optimal solution. 
In EGT, heterogeneity can be related to different aspects of a game, e.g. incentives, interaction topology, learning rates and dynamics. Previous investigations, as~\cite{ Szolnoki2019, Takesue2019, Amaral2018, Szolnoki2018, Zhou2018, Qin2017a, Mann2016, perc_pre08}, showed that heterogeneity can support cooperative behaviours in many competitive scenarios. At the same time, other studies as~\cite{gracia-lazaro_pnas12}, reported small experimental differences between results achieved on heterogeneous and homogeneous networks, when humans play the Prisoner's Dilemma. As another example, findings reported in~\cite{Javarone2017} shows a relationship between the group size of a community and its heterogeneity. This rich spectrum of results highlights the need to further study this subject and to fully understand its range of effects.

As observed in~\cite{amaral_javarone_2020}, perturbations on the payoff matrix of dilemma games actually are responsible for a form of population heterogeneity, as they provide players with a diversified perception of risks and rewards.
For this reason our investigation, while focused on clarifying the role of perturbations in social dilemmas, can provide findings of potential interest for the debate on the relationships between heterogeneity and cooperation in evolutionary games.
Also, the relevance of studying the effects of an heterogeneous risk perception is particularly motivated by a simple fact of everyday life, e.g. individuals facing the same situation will have different risk perceptions~\cite{ Perc2017}. 
Financial trading, emergency medicine~\cite{lawton01}, and even poker games~\cite{javarone_poker_a}, are just a few examples we can mention to appreciate the relevance of an heterogeneity in risk and reward perception in a system. For instance, rational, but risky decisions can be essential to achieve favourable outcomes, and individuals actually show a variety of behaviours in the above listed activities.  
The generality of these considerations makes the payoff values suitable for being considered as stochastic variables, instead as fixed values (as per the classic EGT approach).

One way to study the influence of such random variations is to represent each different environmental condition as a new factor in the equations of a model. Following this method, many authors have done important advances in the understanding of how a number of conditions can drive the system dynamics, such as resource heterogeneity~\cite{Vicens2018}, different behaviours~\cite{Fang2019, Liu2019}, seasonal variations~\cite{Szolnoki2019}, diverse learning rates~\cite{Deng2018}, different death rates~\cite{Junior2019}, interaction topologies~\cite{Szabo2007}, and so on. However, another way to understand these phenomena is to study the behaviour of a population whose evolution can be affected by a payoff matrix constantly perturbed by stochastic noise with zero mean value~\cite{Su2019, Zhang2013, perc_njp11}, irregardless of its origin. 
In other words, as the environmental perturbations are very diverse and frequent, we can suppose that the sum of infinitely many small perturbations acts as a stochastic perturbation around an average value. Indeed, the central limit theorem indicates that the sum of all these uncorrelated perturbations would probably behave as a Gaussian noise. In this scenario, we are not interested in every single source of perturbation, but rather, in their collective effect.

We deem relevant to mention previous seminal works as~\cite{Perc2006, Perc2006a, Perc2006b, tanimoto_pre07b, hofbauer_jet07}, that (to the best of our knowledge) first studied such disorder in EGT. More recently, the effect of payoff noise on phase transitions has been the core of many researches that shed light on how different types of perturbations can lead to the emergence of cooperation~\cite{amaral_javarone_2020, Szolnoki2019, Su2019, Stollmeier2018, Alam2018, Hilbe2018, Javarone2016b, Tanimoto2016, Yakushkina2015, wang_z_pre14b, Zhang2013, anh_tpb12}. So, here, starting from the model proposed in~\cite{amaral_javarone_2020}, which introduces payoff perturbation in the context of the imitation update rule for all payoff entries, we perform a full analysis under different settings, and then we study the microscopical mechanism that leads to the observed macroscopic results.

After introducing the proposed model in Section~\ref{sec:Model}, we show the results of numerical simulations (i.e., Section~\ref{sec:Results}) and, eventually, in Section~\ref{sec:Conclusions}, we discuss main findings and their possible implications.

\section{Model}
\label{sec:Model}
For the sake of simplicity, we consider two-strategy games, where players can either cooperate (C) or defect (D). Mutual cooperation yields a payoff $R$ (reward) and mutual defection yields $P$ (punishment). A defector receives a payoff equal to $T$ (temptation) when interacting with a cooperator, that in turn receives a payoff equal to $S$ (which usually stands for Sucker's payoff). Then, we introduce the effects of small payoff perturbations, say $\varepsilon$, randomly occurring on the off-diagonal elements of the payoff interaction matrix.
As result, the payoff matrix of any two-player interaction can be written as follows
\[
 \bordermatrix{~ & C & D \cr
                  C & R & S+ \varepsilon _S \cr
                  D & T+ \varepsilon _T & P \cr},\]

\noindent where $T\in[0,2]$ and $S\in[-1,1]$. Without loss of generality, we set $R=1$ and $P=0$. A similar approach was also introduced in~\cite{Perc2006a}.
It is worth clarifying that $\varepsilon_T$ and $\varepsilon_S$ are independent random variables with zero average value ($\bar{\varepsilon}=0$), drawn from a uniform distribution unless stated otherwise. At each new player interaction, the fluctuation is randomly drawn again, meaning that the noise is not cumulative, nor fixed in time or space. This is similar to the annealed disorder case of condensed matter physics~\cite{Tanimoto2016, Fiore2018}.

\begin{figure}
  \includegraphics[width=6cm]{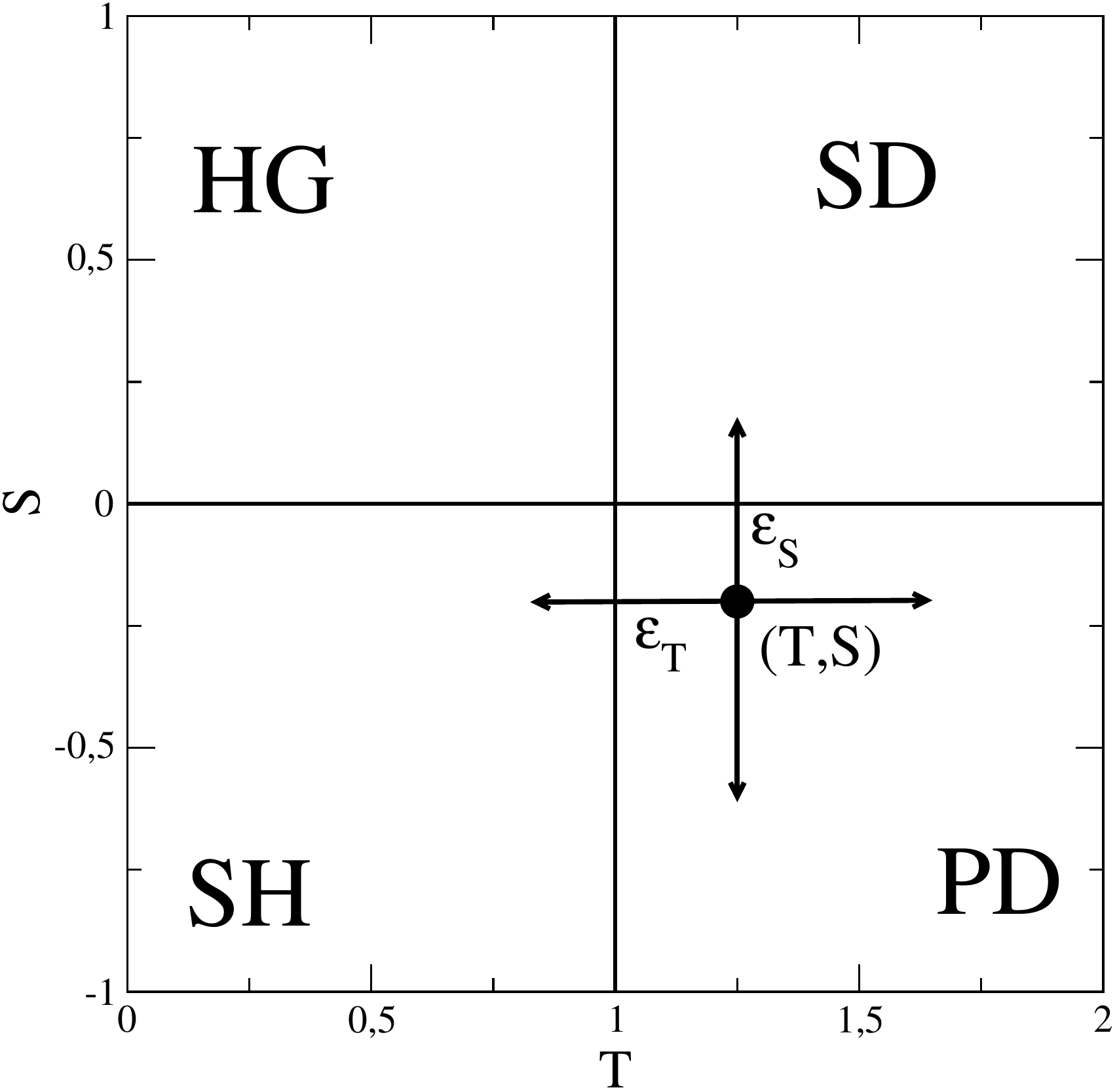}
  \caption{$T\times S$ parameter space with $R=1,P=0$, spanning four classes of games. The payoff fluctuation acts over $T$ and $S$ simultaneously and uncorrelated. Note that local fluctuations can lead players to play different classes of games depending on the fluctuation strength.}
  \label{diagram}
\end{figure}

Here, the control parameter for the proposed model is the perturbation strength $D$, where for a uniform distribution we set $-D<\varepsilon<D$. It is useful to remark that this payoff matrix parametrization spans four different classes of games in the $\{T,S\}$ parameter space: prisoner's dilemma (PD), snow-drift (SD), stag-hunt (SH), and harmony games (HG)~\cite{Szabo2007, perc_bs10} ---see Figure~\ref{diagram}. We emphasize that even if the average value of the perturbation is zero, it can be able to (locally) change the game class to a more friendly or more competitive environment from time to time. However, in the long run, there should be, on average, no unilateral contribution to either.

A general core aspect of evolutionary games is given by the strategy update dynamics. We can understand it as a two-step process where: $(1)$ players interact with their neighbours accumulating a payoff, then $(2)$ they may change strategy according to an update rule. Such rules can be defined by taking into account different aspects of any given system. Here we consider the usual Imitative rule \cite{Szabo2007}, the Ising (or Glauber Dynamics) rule \cite{Amaral2017}, and the Dynamic Win-Stay-Lose-Shift (WSLS) rule \cite{Amaral2016b}. This is done so we can study the robustness of the effects created by payoff perturbation since previous works have extensively shown how the update rules can lead to different behaviours~\cite{Takesue2019, Fang2019, Szolnoki2018, Danku2018, Amaral2018, Dercole2019}.

We first present the most usual update rule, the imitative dynamic. Accordingly, one player, say $i$, updates its strategy by comparing its payoff with that of one randomly chosen neighbour, say $j$. Player $i$ adopts the strategy of player $j$ with probability
\begin{equation}\label{imitateeq}
p(\Delta u_{ij})=\frac{1}{ 1+e^{-(u_{j}-u_{i})/k} },
\end{equation}
\noindent where $k$ is the irrationality level in the decision process~\cite{Szabo2007} (similar to the temperature in statistical physics), while $u_{i}$ and $u_{j}$ represent the player $i$'s and $j$'s payoff. We use $k=0.1$ for all simulations unless stated otherwise. The imitation rule is a non-innovative dynamic~\cite{Szabo2007,Nowak2006}, because a player can only change its strategy selecting among those available in the population. This is one of the most explored rules in EGT, and it is associated with the replicator dynamics observed in biological evolution~\cite{Smith82}. Because of the extensive research on this setup, we will focus our analysis mainly on this model.

We present next the so-called Ising rule (also known as Glauber Dynamics \cite{Glauber1963}), where players change strategy with probability
\begin{equation}\label{isingeq}
p(\Delta u_{i})=\frac{1}{ 1+e^{-(u_{i}^*-u_{i})/k} }
\end{equation}
\noindent with $u_{i}$ player $i$'s payoff, and $u_{i}^*$ its potential payoff at a next iteration, if it changed strategy while everything else (i.e. the strategy set of its neighborhood) remained the same. Recently, the Ising rule has been studied in~\cite{Amaral2017, Amaral2018, szabo_jtb12b}, showing that it leads to very different dynamics when compared to imitation models.
In mathematical terms, this setting is equivalent to a Monte-Carlo process used to describe the dynamics of spins in a Heisenberg model~\cite{binder_prb80}. In the context of EGT, this update rule is regarded as a player asking himself what would be the benefits of changing its strategy to a different one. This is closely related to rational analysis of a situation (see also~\cite{Javarone2016d} on this topic), instead of the reproduction of the 'fittest' behaviour. 

Lastly, we also implement the Win-Stay-Lose-Shift (WSLS) update rule with dynamic aspiration~\cite{Amaral2016b}. The WSLS strategy relies on cognitive capabilities, instead of replicating process~\cite{bonawitz_cg14, perc_pone11, pacheco_prl06, chen_jsm15}. In this case, players change strategy depending on the degree of satisfaction with their current payoff, in comparison to the average payoff of their neighbourhood. 
The probability of a chosen player to change its strategy to the opposite one is given by:
\begin{equation}\label{wslseq}
p(\Delta u_{i})=\frac{1}{ 1+e^{-( \bar{u}-u_{i})/k} },
\end{equation}
\noindent where $\bar{u}$ is the average payoff of the player $i$'s neighborhood.

To implement such dynamics we use an asynchronous Monte-Carlo protocol, so that a random player, say the $i$-th, is selected and its cumulative payoff, as well as those of its first and second degree neighbours, are calculated. Then the $i$-th player can change its strategy according to the defined update rule. One Monte Carlo step (MCS) consists of this process being repeated until each player has had the opportunity to change its strategy (that is, $N$ times). So, we wait for the system to reach a dynamic equilibrium (around $10^4$ MCS's), and then we average the values over the final $1000$ steps. This is repeated for $50-100$ different samples with the same parameters. A square lattice, with Von Neumann neighbourhood and periodic boundary conditions with $N=10^{4}$ individuals, is the background for our simulations. A detailed discussion on Monte Carlo methods in evolutionary dynamics is provided in~\cite{landau_00, Szabo2007}.

\section{Results}
\label{sec:Results}
\begin{figure}
  \includegraphics[width=8.5cm]{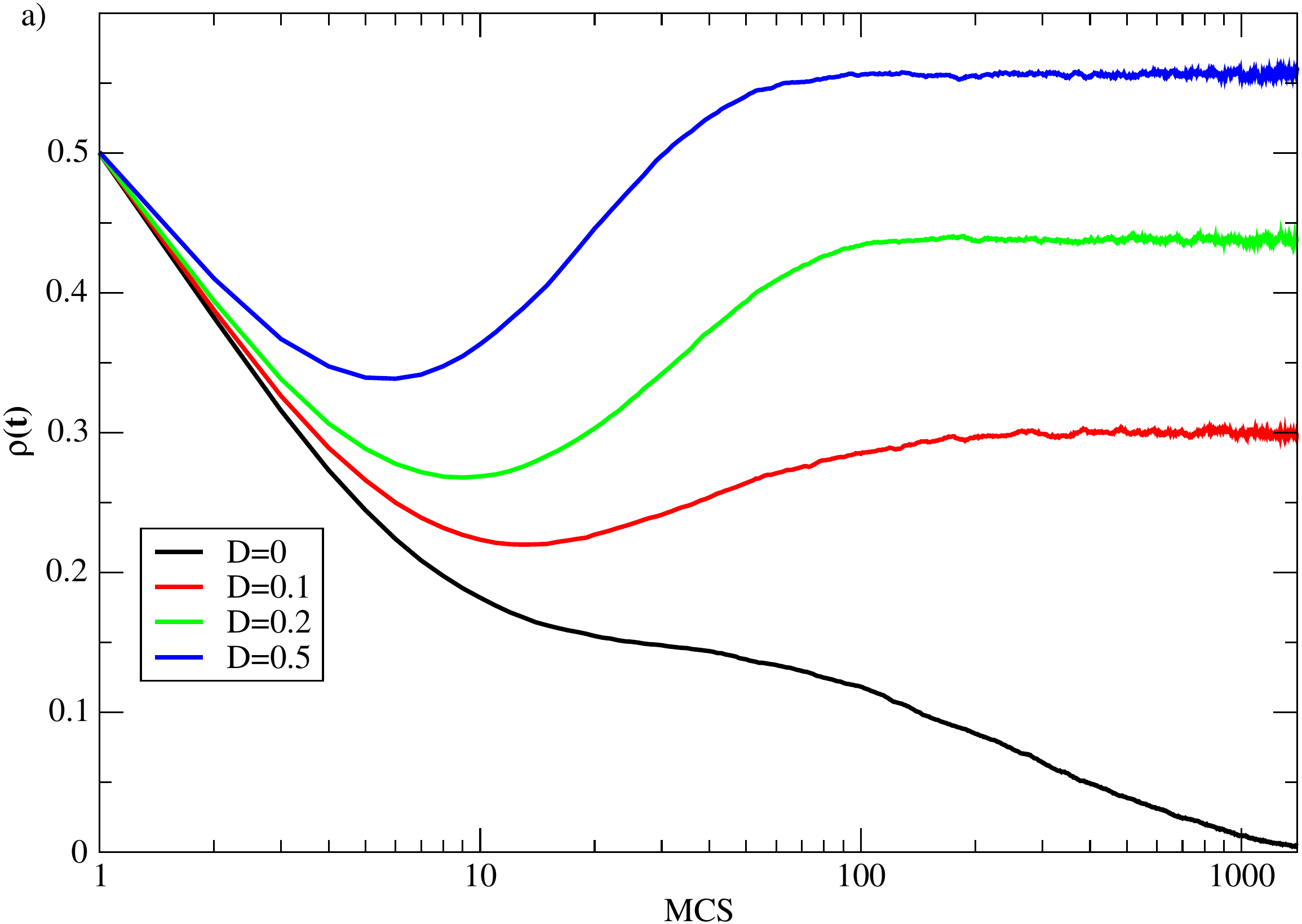}
  \includegraphics[width=8.5cm]{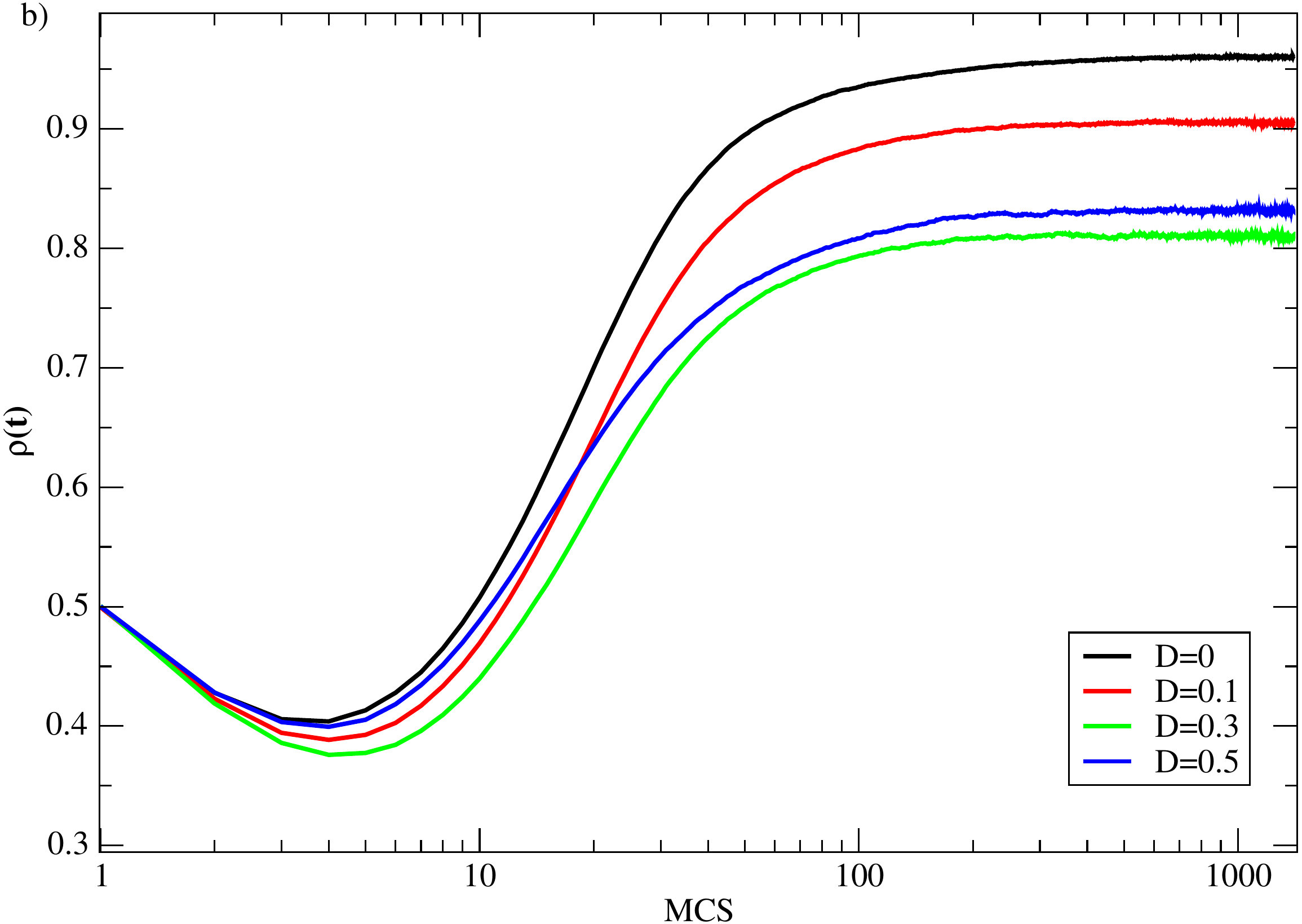}
  \caption{Average evolution of cooperation, $\rho(t)$, for for $100$ samples using the imitative update rule. We set $T_c=1.04$ in a), the critical extinction point for cooperation. Even when the noise is around $5\%$ of the $T$ value, cooperation can re-emerge. Figure b) presents similar analysis for $T=0.96$. Notice that the $x$ axis is logarithmic. }
  \label{timeaverage}
\end{figure}

In this section, we show the results of numerical simulations of the proposed model. For the sake of clarity, irrelevant error bars are hidden. Figure~\ref{timeaverage} a) presents the average cooperation fraction, $\rho(t)$, of a population as time passes, considering the imitative rule. Also, it shows the evolution of cooperation for the parameters $T_c=1.04$ and $S=0$. We remind that for the imitative model with no payoff perturbation, $T_c=1.04$ corresponds to the phase transition point where cooperation becomes extinct. This was chosen to highlight the positive effect of payoff perturbation, since its strongest effect happens near the phase transition boundary. Remarkably, adding a perturbation as low as $D=0.05$ can prevent said extinction, and for $D=0.2$, cooperation can increase even up to $40\%$. We note that this increase in cooperation also happens for a wide range of $T$ values. Qualitatively similar results were obtained for the Ising and WSLS update rule regarding the temporal evolution of the population.

The analysis of the temporal evolution indicates that the noise affects almost immediately the cooperative behaviour. Using the definitions of Tanimoto~\cite{Tanimoto2016}, we can observe that the ``Endurance'' phase (where cooperation initially falls) is drastically affected. Notably, even for $MCS<20$, the noise enhances cooperation preventing its extinction. The ``Expanding'' phase is therefore reached for very short times, allowing cooperation to flourish. Also note that Figure~\ref{timeaverage} b) presents the average temporal behavior for $T=0.96$. We observe that the perturbation can be detrimental to cooperation if $T<1$, that is, for more fraternal games. Note however that this detrimental effect is weaker than the positive effect when $T>1$.

\begin{figure}
  \includegraphics[width=8.5cm]{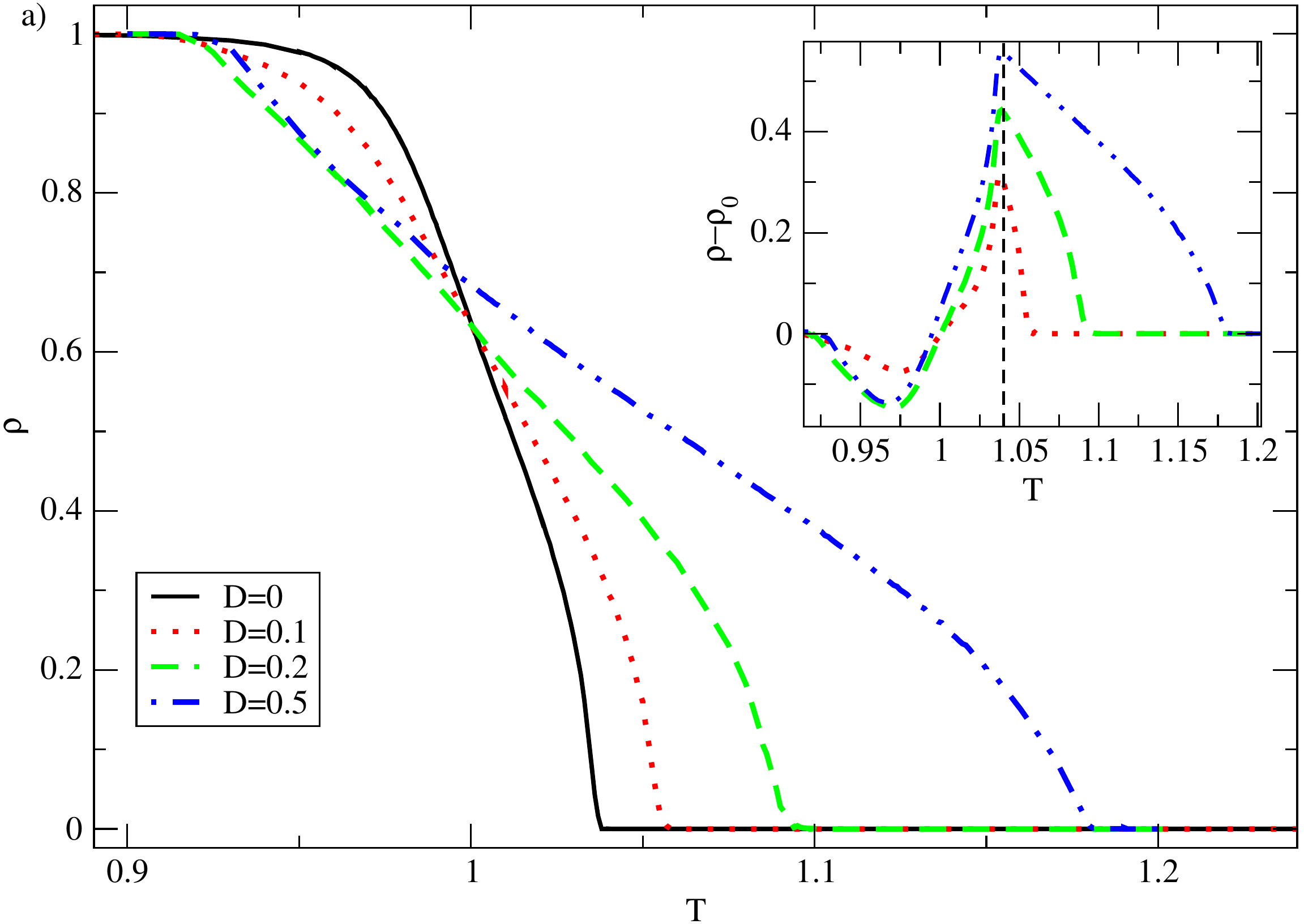}
  \includegraphics[width=8.5cm]{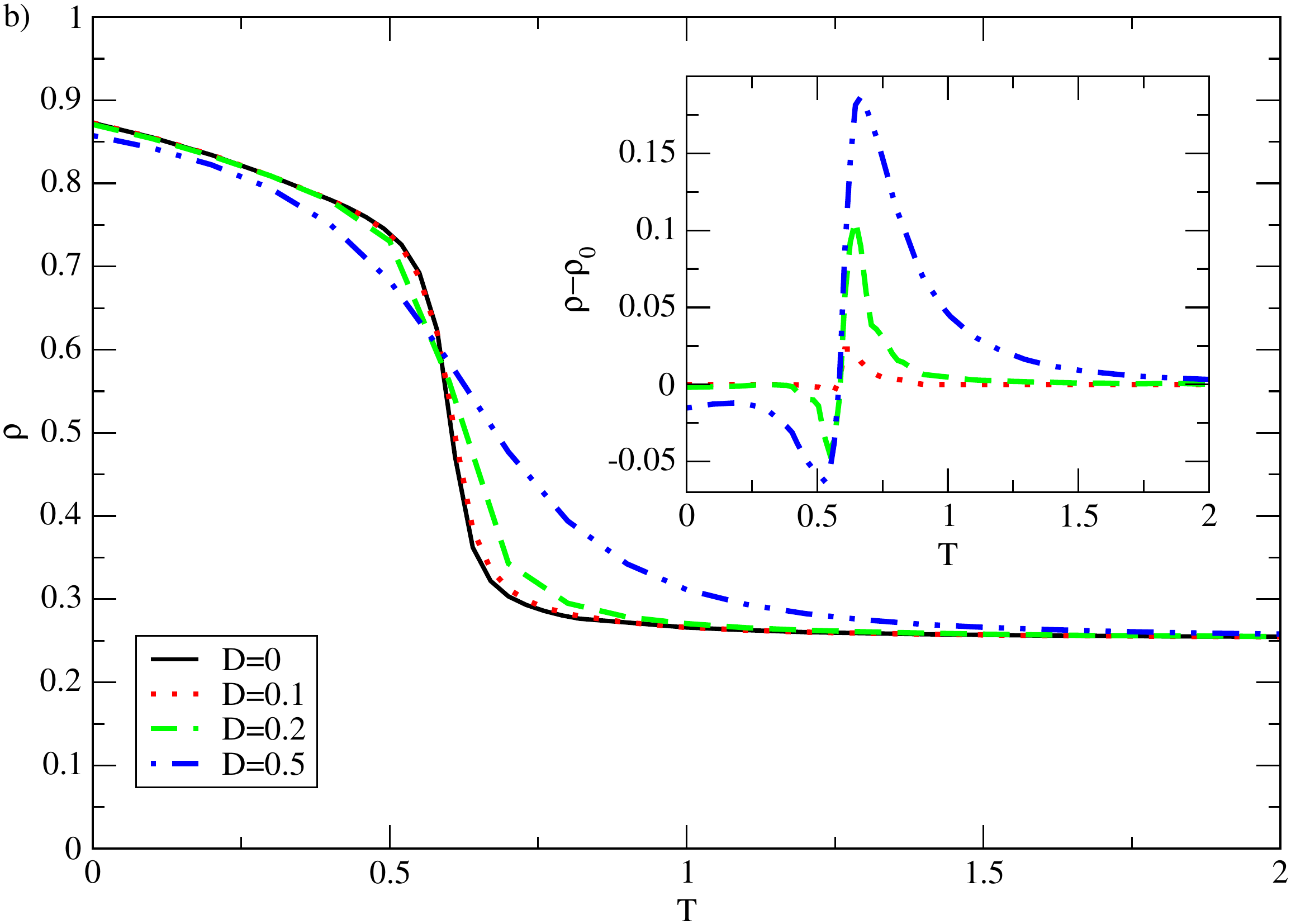}
  \includegraphics[width=8.5cm]{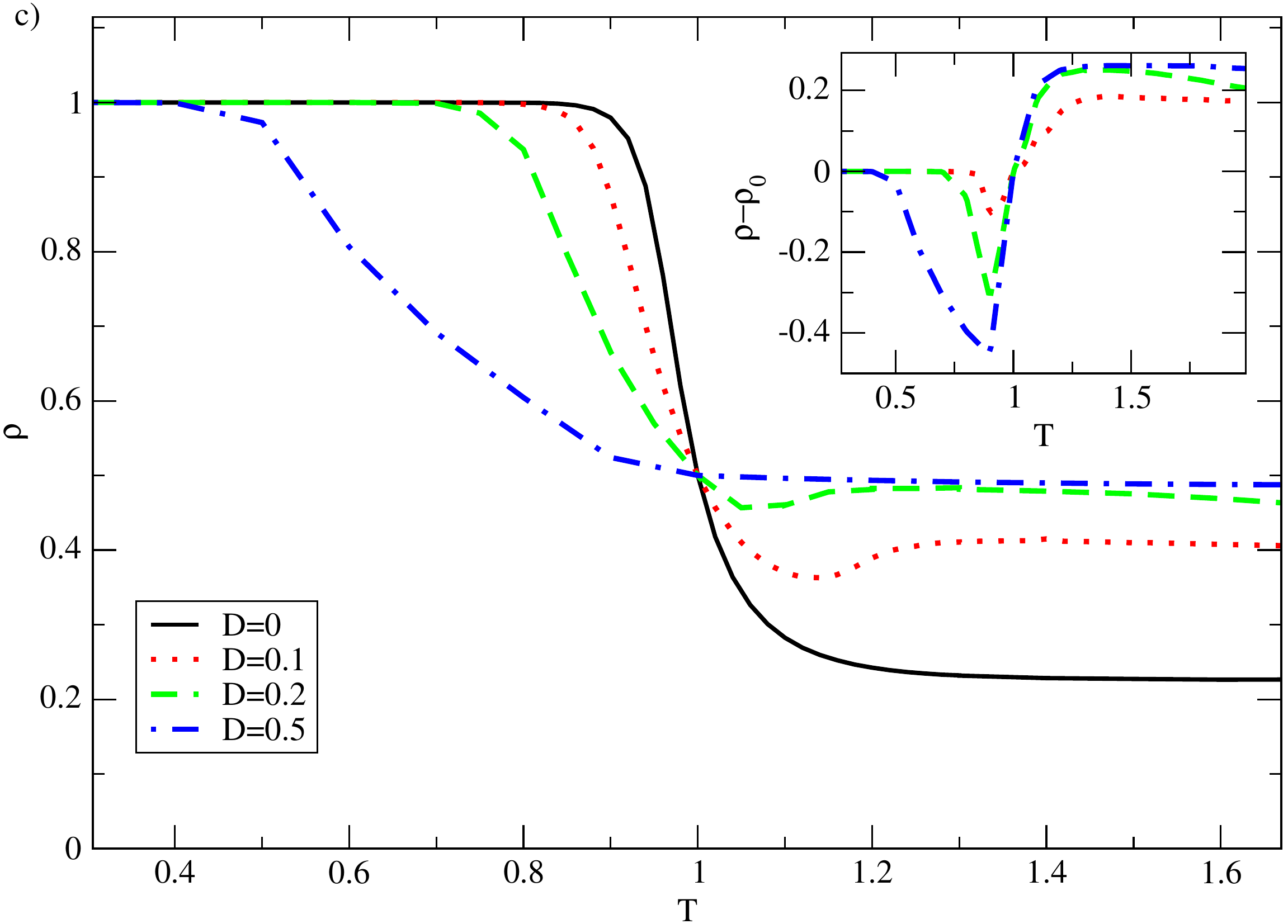}
  \caption{Average final cooperation level ($\rho$) as a function of $T$ for the imitative (a), WSLS (b) and  Ising(c) models using the uniform noise distribution. The inset shows $\rho-\rho_0$, where $\rho_0$ is the cooperation value for the case $D=0$.}
  \label{Tvar_uniform}
\end{figure}

Next, we focus on the final equilibrium fraction of cooperators, $\rho$, for the whole range of $T$. For simplicity, we present results for the weak prisoner's dilemma configuration (i.e. $S=0$). We ran simulations for different $S$ values, and the general trends are maintained. Figure~\ref{Tvar_uniform} presents the outcomes achieved by using the uniform noise distribution in all considered update rules (i.e. Imitative, WSLS and Ising). Its inset shows $(\rho-\rho_0)$, where $\rho_0$ is the cooperation value for the case $D=0$. This is especially useful to compare the effects of the payoff perturbation with the unperturbed case, filtering off the effects of varying $T$.
It is interesting to note that there is a general trend in all models, i.e. the noise increases cooperation in the PD region ($T>1$), and such enhancement grows with the noise amplitude, $D$. We stress here how the three update rules have very different dynamics, and that even so, the general effect of the payoff noise was maintained in all three cases. 
The inset allows understanding the level of enhancement caused solely by the noise. In particular, the Imitative and WSLS rules present a clear peak, that grows with $D$ while maintains its position (relative to $T$) regardless of the noise level. 
It is also very interesting to note that this peak happens in $T_c=1.04$ for the imitative model, and after that, the positive effect begins to decline. This again reinforces that the benefit of payoff perturbation is most strong near the phase transition. At the same time, for $T<1$, the noise can dampen cooperation in different manners for each case. Using the imitative updating rule, the related drop is very shallow ($~15\%$) and almost independent of $D$. That is similar to results achieved by using the WSLS rule. On the other hand, the Ising rule presents a strong drop in $(\rho-\rho_0)$, which is very dependent on $D$.

To observe how general the effect can be, we also analysed a similar setting but with a Gaussian distribution, instead of a uniform one. In this case, we set our control parameter as $D=\sigma$ (i.e. the standard deviation of the distribution). Let us stress that we cannot directly compare the control parameter $D$ for both cases since, in the uniform distribution, $D$ is the range of the distribution, whereas in the Gaussian case, $D$ is the standard deviation. Even so, $D$ relates to the perturbation strength in both scenarios. The results are very similar for the three update rules, with minor quantitative differences. The general difference being that for Gaussian distributions, cooperation is more enhanced than for uniform ones given the same value of $D$. This can be credited to the fact that in a Gaussian distribution with $\sigma=D$, although most perturbations will be equal or smaller than $D$, there will be rare stronger perturbations occasionally. This, in turn, translates to a more noisy system that can strongly enhance cooperation. 

\begin{figure}
  \includegraphics[width=8.5cm]{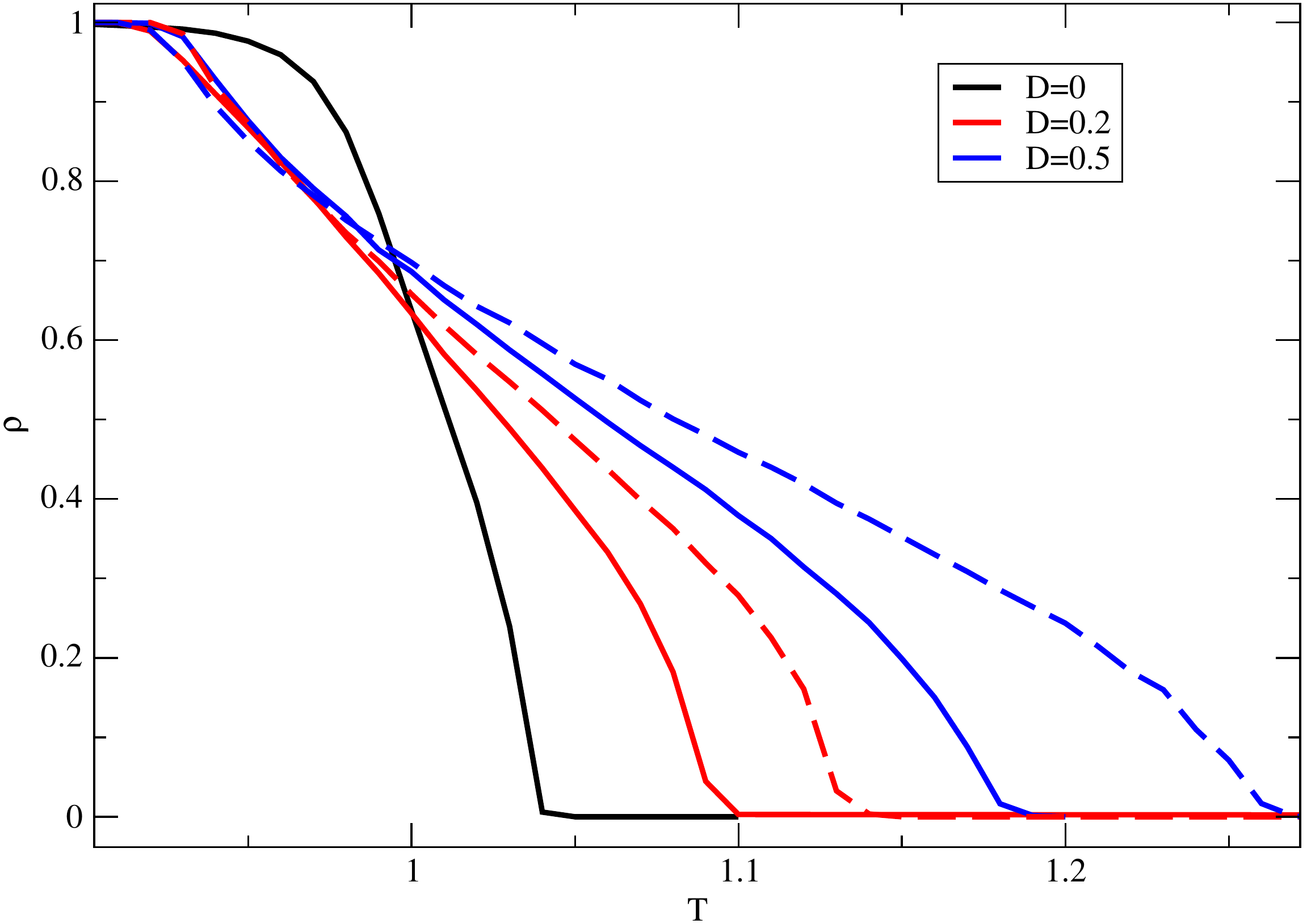}
  \caption{Average final cooperation level as a function of $T$ for the imitative model, comparing the uniform (continuous lines) and Gaussian (dashed lines) distribution. Here $D$ stands for the strength of the perturbation. In the uniform distribution $D$ is the range of the perturbation while in the normal distribution it is the standard deviation.}
  \label{imitcompare}
\end{figure}

In figure~\ref{imitcompare} we present a comparison between the uniform and Gaussian distributions, using the imitative update rule. For the sake of clarity, we present only $D=0.2$ and $D=0.5$, but the effect was similar for all range of $D\in [0,1]$. As expected (assuming the same control parameter $D$ for both cases), the Gaussian distribution has a more pronounced effect. This suggests that the cooperation enhancement stems from rare and influential fluctuations, that are more common in the Gaussian distribution. This is in accordance with a similar hypothesis presented in \cite{Perc2006, Perc2006a, Perc2006b}.

\begin{figure}
   \includegraphics[width=8.5cm]{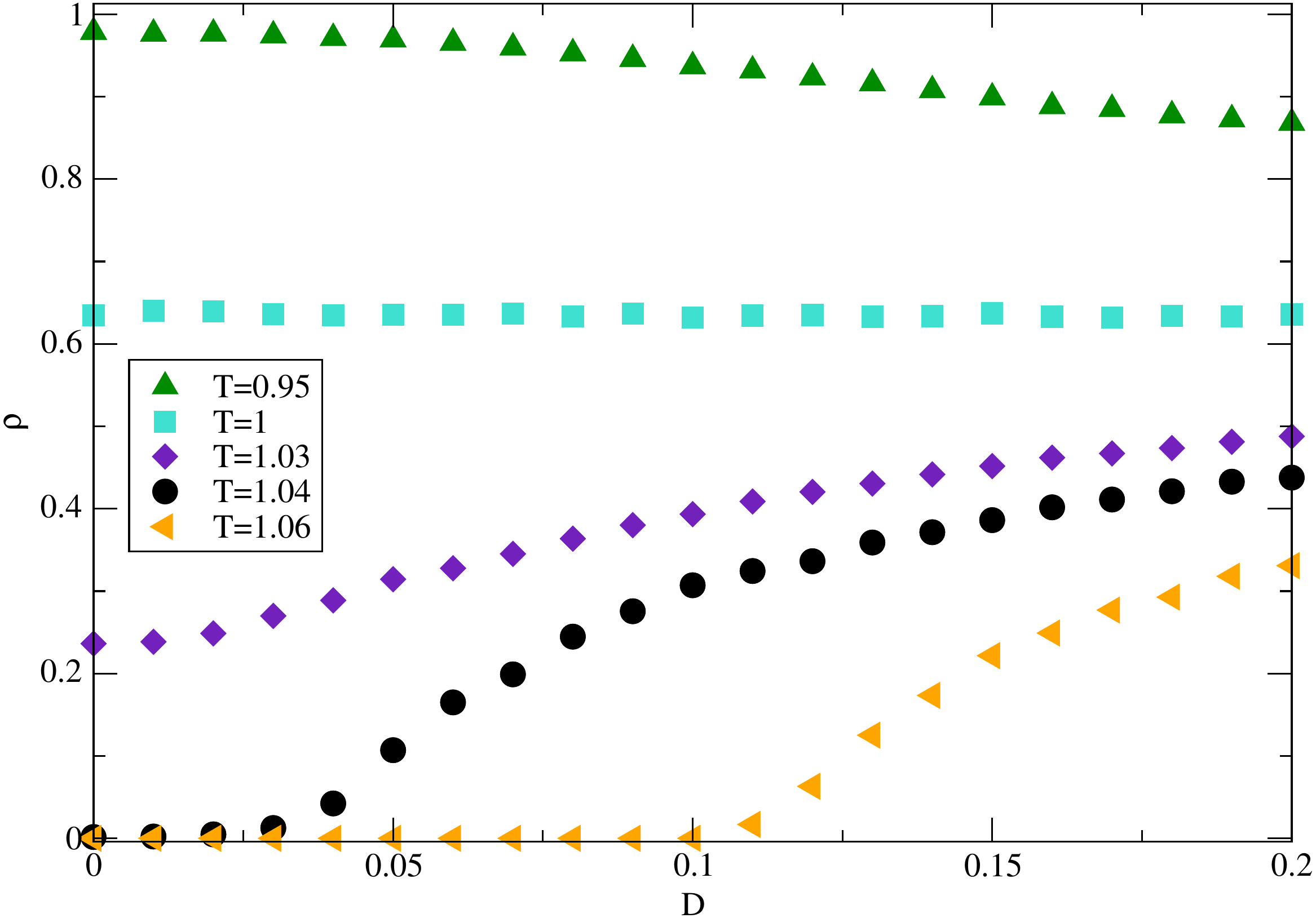}
  \caption{Average final cooperation level as a function of perturbation strength $D$ for different values of $T$. Black circles represent the critical point $T_c=1.04$, were the perturbation effect is stronger.}
  \label{vardelta}
\end{figure}

Considering the imitative rule, we varied the noise level, $D$, for different $T$ values. Results are presented in Figure~\ref{vardelta}. The perturbation has a positive and continuous effect on the cooperation in regions where $T>1$, and a (small) detrimental effect if $T<1$. Specifically, the strongest cooperation enhancement happens exactly at the phase transition ($T_c=1.04$). We see that payoff perturbation enhances cooperation for egotistic games ($T>1$) while it can dampen said cooperation for fraternal games ($T<1$).

\begin{figure}
   \includegraphics[width=8.5cm]{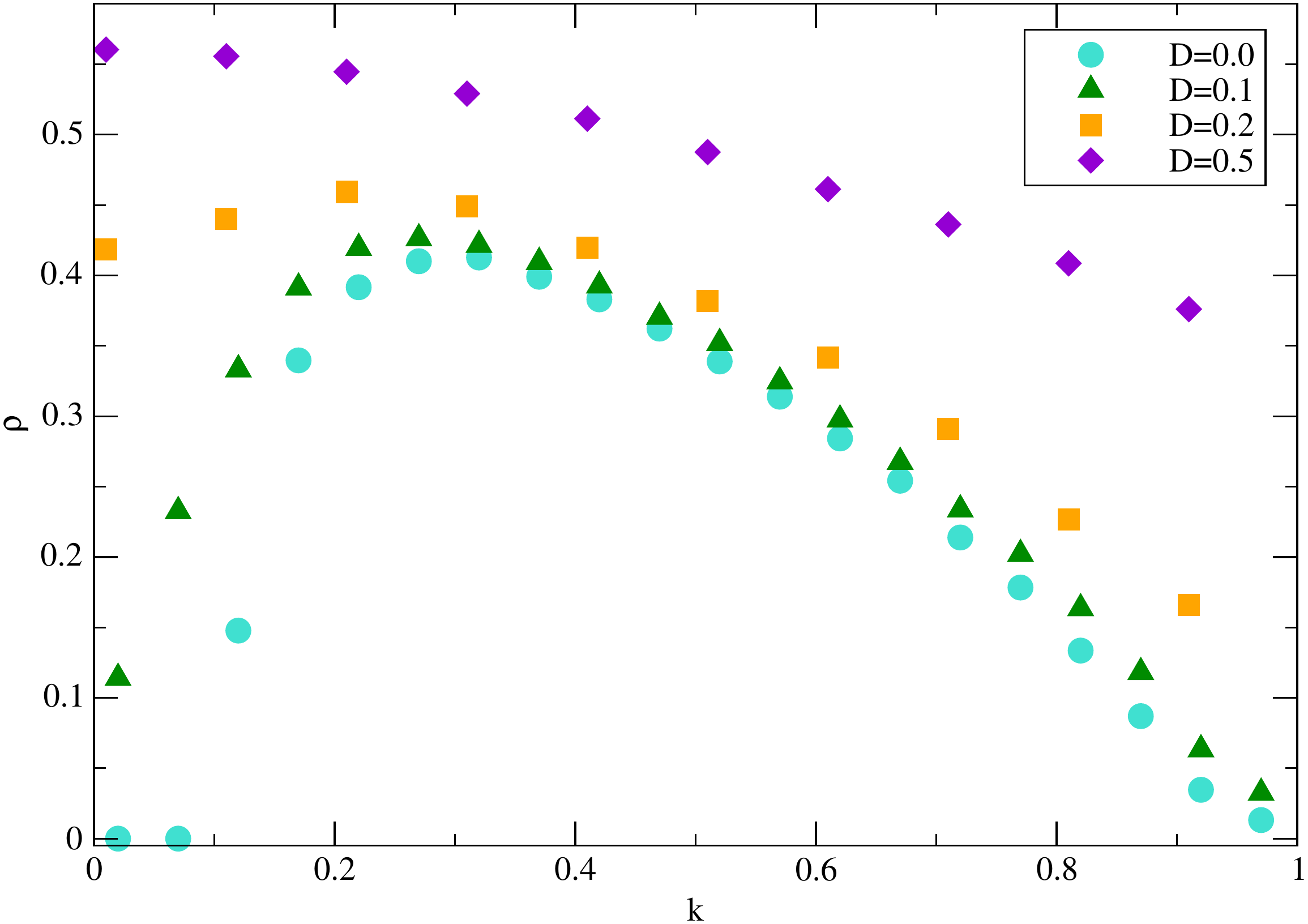}
   \caption{Average final cooperation level as a function of irrationality, $k$, in the phase transition $T=1.04$ for the imitative model. As expected, there is an optimal irrationality level for $D=0$. Even so, payoff fluctuations always increases cooperation for the whole range of explored $k$ values.}
  \label{rho_k_D}
\end{figure}

To understand whether the cooperation boost is caused by any kind of random fluctuation, we compare the effects of payoff perturbation with the noise generated by high irrationality values, $k$, ~\cite{Javarone2016d, vukov_pre06}. As demonstrated by Szabó et al \cite{szabo_pre05}, irrationality can have beneficial effects on the maintenance of cooperation for some parameter regions. Nevertheless, this is not a linear effect, and there is an ``optimal'' irrationality level, after which the system starts to behave randomly, destroying cooperation. Figure~\ref{rho_k_D} illustrates the final average cooperation level as a function of irrationality $k$ for different payoff perturbation strengths. We see that the payoff perturbation boosts cooperation regardless of the irrationality level. Even more, we see that the cooperation enhancement granted by intermediate levels of irrationality is additive with the beneficial effect of the payoff perturbation (more strongly present in the region $0.15 < k < 0.6$). We used $T=1.04$ for the figure, but this effect remained similar for different $T$ values and different update rules in our simulations.

We proceed to analyse which microscopic mechanism is responsible for the cooperation enhancement due to payoff perturbation. It is worth noting that lattice snapshots were not useful in this regard in all three update rule settings. The population has the same general spatial distribution with and without perturbation, i.e., in both cases, cooperators form the usual clusters surrounded by a sea of defectors. The main difference being only that the total fraction of cooperators is greater as we include perturbations. While this does not present the cause of the cooperation enhancement, we see that it is not directly related to a strong spatial distribution effect.

We stress that while the perturbations have zero average value, they can locally change the game class being played at each round. Based on this fact, we analyse how players (locally) fluctuate to a more fraternal or egotistic game. To do so, we define for each player, at each Monte-Carlo step, the variable $\phi=\varepsilon_S-\varepsilon_T$. If $\phi>0$ ($\phi<0$) the game being played at a given time will be more fraternal (egotistic). Note that for every player, $\bar{\phi}=0$ for long times. Next, we obtain at each time step the population fraction of cooperators and defectors that had $\phi>0$ in the previous round (named fraternal cooperators, $C_F$, and fraternal defectors, $D_F$). We do the same for cooperators and defectors with $\phi<0$ (named egotistic cooperators, $C_E$, and egotistic defectors, $D_E$). By doing so, we can understand if players strategies are correlated with $\phi$ even if $\varepsilon$ is randomly drawn at every interaction. 

\begin{figure}
   \includegraphics[width=7cm]{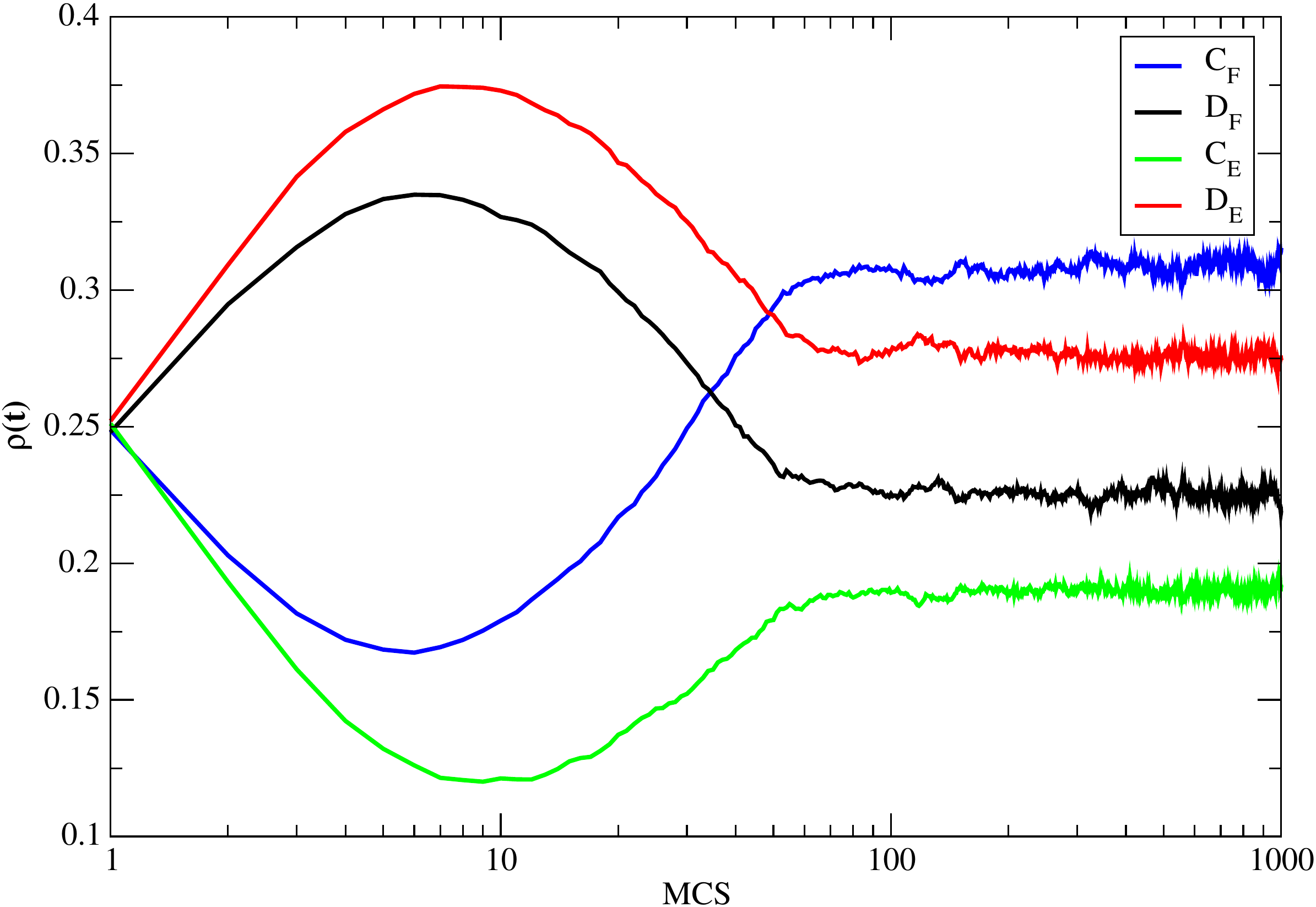}
  \caption{Average fraction of each sub-population over 100 Monte-Carlos runs. Here we used $D=0.3$ and $T=1.04$, where cooperation would be extinct if not for the payoff perturbation. Fraternal cooperators ($C_F$) quickly dominate the population being followed by Egotistic Defectors ($D_E$), Fraternal defectors ($D_F$) and lastly  Egotistic Cooperators ($C_E$).}
  \label{flux}
\end{figure}

Figure \ref{flux} presents the average evolution of the four sub-populations for the imitative model in the region where the perturbation has its strongest effect, i.e. $T_c=1.04$. The main effect of payoff perturbation is to separate the sub-populations of players who play, on average, more fraternal or egotistic games. Specifically, fraternal cooperators ($C_F$) grows more than any other sub-population, being followed by the egotistic defectors ($D_E$). It is important to note however that even if the perturbations are symmetric, $C_F>D_E$.

\begin{figure}
    \includegraphics[width=7cm]{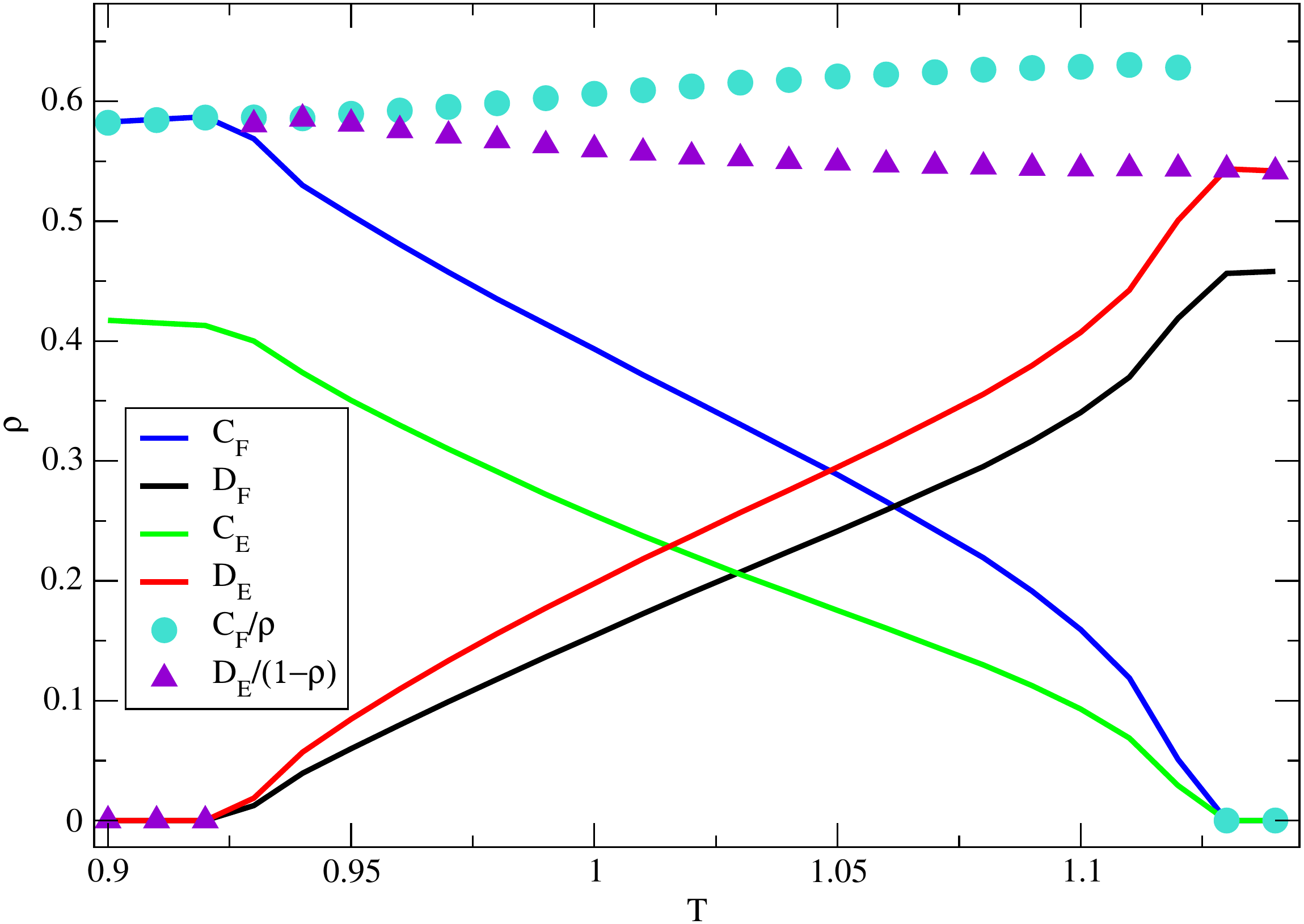}
  \caption{ Average final fraction of each sub-population (lines) as we vary $T$. Here we used $D=0.3$. Symbols presents the average normalised fraction of cooperators that are fraternal, i.e. $C_F/\rho$, and the average fraction of defectors that are egotistic, i.e. $D_E/(1-\rho)$. Sites that play more fraternal games on average will tend to become cooperators at a higher frequency then egotistic sites will become defectors.}
  \label{fluxtransi}
\end{figure}

To further explore this, we obtained the fraction of all sub-populations for different $T$ values. The results are shown in Figure~\ref{fluxtransi}. In it, we can see that this effect happens for all the relevant $T$ range. On average, sites that have fluctuations leading to $\phi>0$ will tend to become cooperators with a higher frequency than sites that had $\phi<0$ will become defectors. Note that the symbols in figure~\ref{fluxtransi} present the (normalised) fraction of cooperators that are fraternal, i.e. $C_F/\rho$ (and the normalised fraction of defectors that are egotistic, i.e. $D_E/(1-\rho)$). This is  especially important, since as $T$ increases, the total fraction of cooperators decreases, but even so, $C_F/\rho>D_E/(1-\rho)$. In other words, even if total cooperation decreases, the perturbation still induces a flux of cooperators to sites where more fraternal games were played. This, in turn, sustains cooperation for greater ranges of $T$.

\begin{figure}
    \includegraphics[width=7cm]{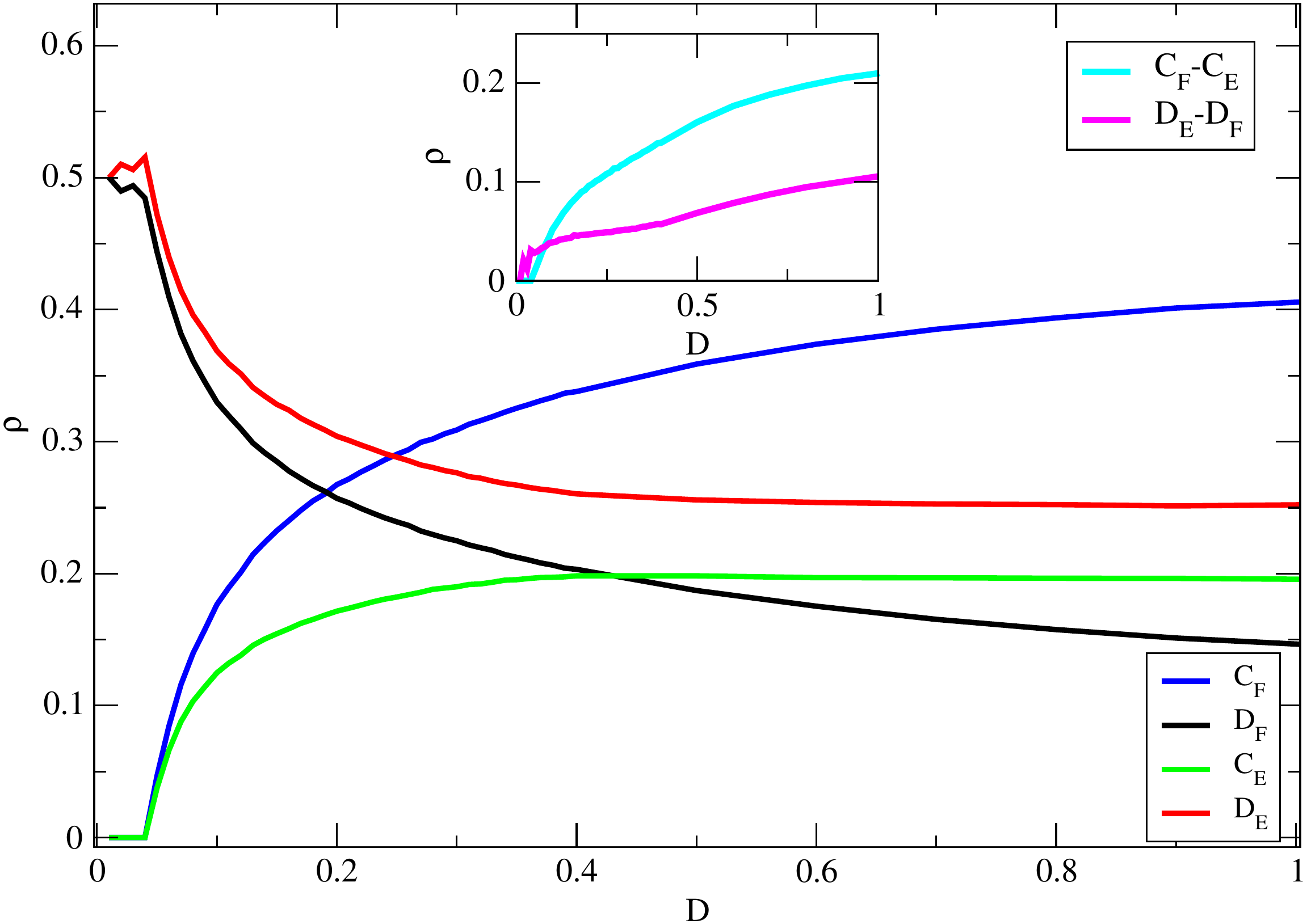}   
  \caption{ Average final fraction of each sub-population (lines) as we vary $D$. Here we used $T=1.04$. The inset shows the difference of the ,average, fraternal and egotistic sub-populations, $(C_F-C_E)$ as well as $(D_E-D_F)$.}
  \label{fluxDtransi}
\end{figure}

This analysis was also performed for different perturbation strength values ($D$) to observe how it can affect the sub-populations. The results are shown in Figure~\ref{fluxDtransi}. As expected, the increase in both sub-populations of cooperators is monotonous with $D$. The inset shows the difference $(C_F-C_E)$ as well as $(D_E-D_F)$. In it, we can see that even if a stronger perturbation can lead to a higher value of $(D_E-D_F)$, this value is always lower than $(C_F-C_E)$. In other words, agents who play more fraternal sites on average will have a greater probability of becoming cooperators. On the other hand, the opposite is not true for players that have played, on average, more egotistic games, as their average fraction will be smaller than that of the cooperators. This is the main mechanism behind cooperation enhancement. Simulations for the WSLS and Ising update rules remained similar with small quantitative differences, showing how robust this phenomenon is.

Let us summarise the main mechanism that can increase cooperation in the proposed model; while players have symmetrical perturbations, defectors locally benefit from a more egotistic game, but cannot maintain said benefit as their clusters do not obtain benefits from the perturbation ($P$ is not perturbed). At the same time, cooperators can form spatial structures that are more robust against negative perturbations, heaping the long term benefits of positive perturbations.  In other words, cooperators are able to form positive feedback loops with the local positive payoff fluctuations, while defectors cannot. This effect is very similar to what was observed in~\cite{Amaral2016, szolnoki_epl14b}, where a very similar positive feedback loop helps the promotion of cooperation for the so called multi-games. Indeed, this seems to be a general property of evolutionary games, i.e., cooperators can benefit from a variety of fluctuations using positive feedback loops, whereas defectors cannot.

\begin{figure}
  \includegraphics[width=5cm]{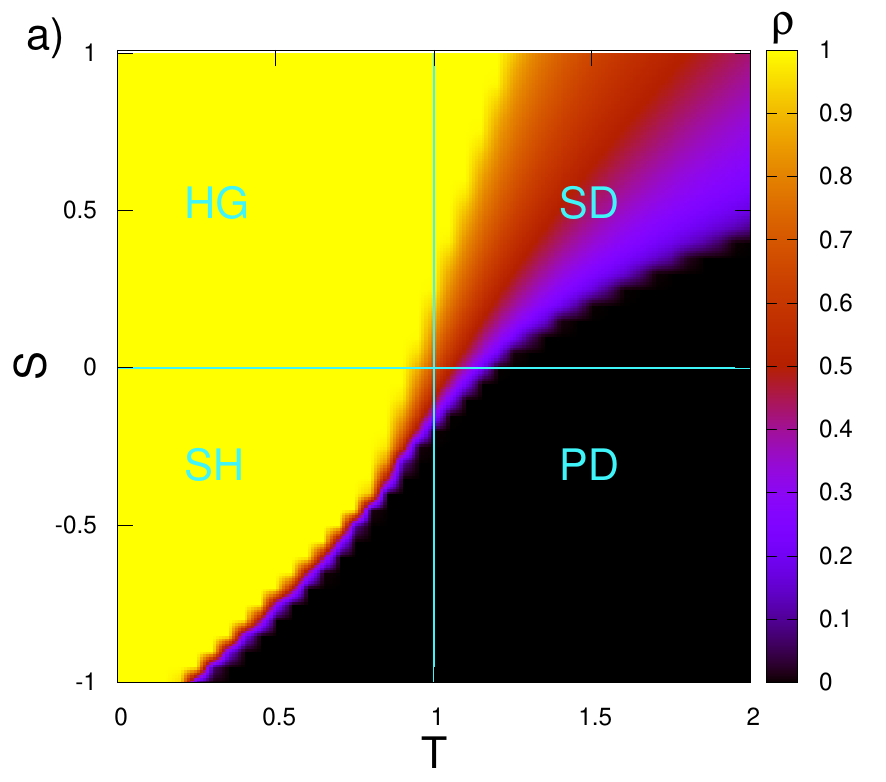}
  \includegraphics[width=5cm]{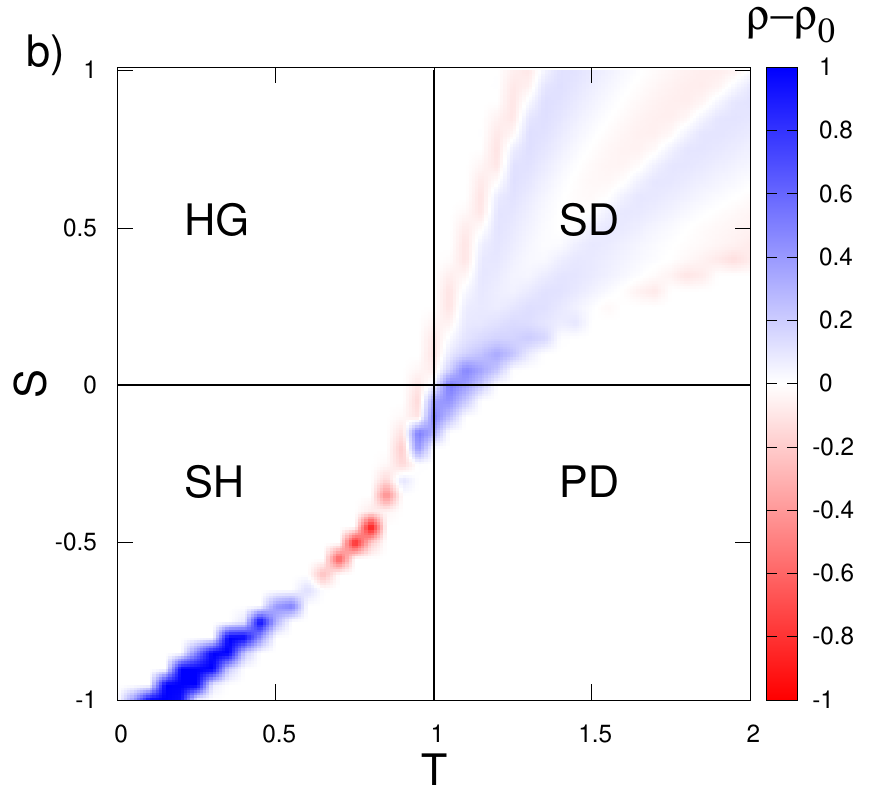}
  \caption{ $T \times S$ parameter space for the imitative model. Final cooperation fraction is represented in colours. a) high perturbation ($D=0.5$) and b) difference $(\rho-\rho_0)$. }
  \label{phaseimit}
\end{figure}

\begin{figure}
  \includegraphics[width=5cm]{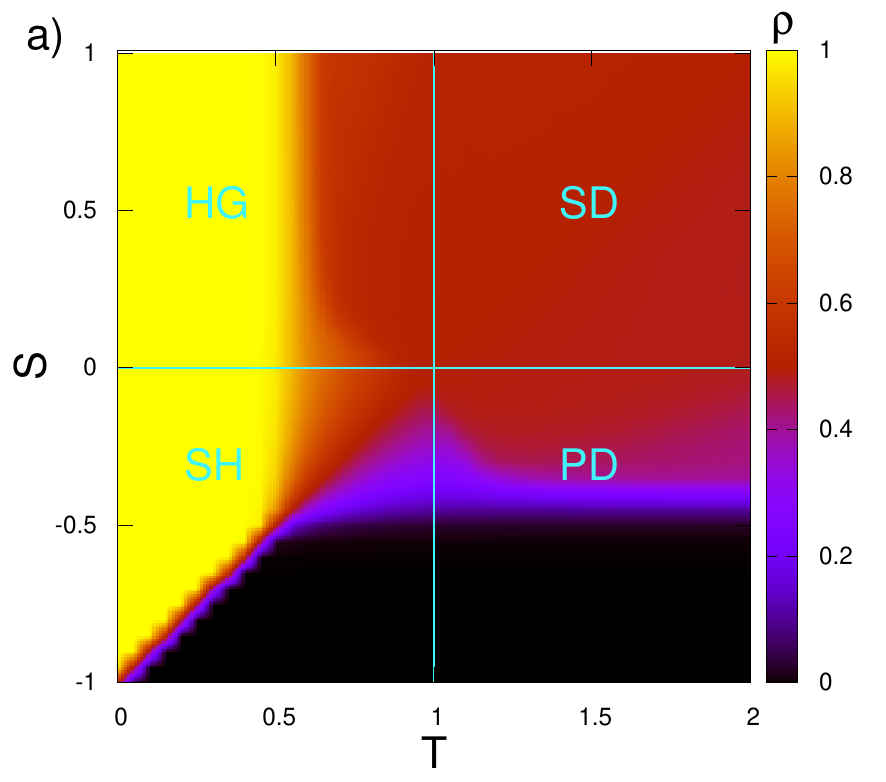}
    \includegraphics[width=5cm]{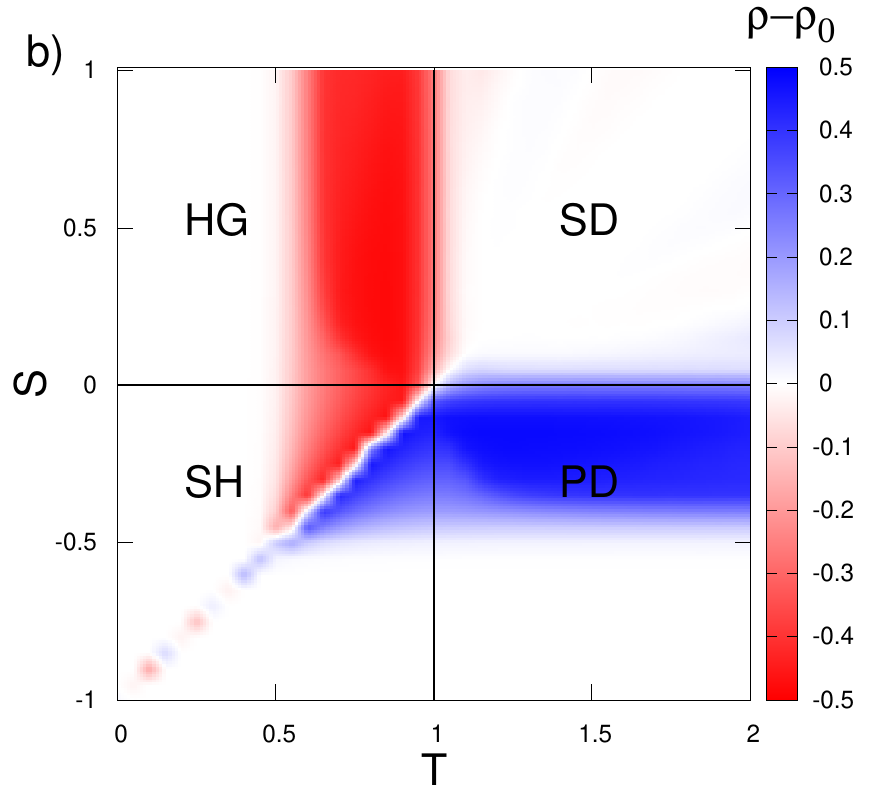}
  \caption{$T \times S$ parameter space for the Ising model. Final cooperation fraction is represented in colours. a) high perturbation ($D=0.5$) and b) difference $(\rho-\rho_0)$.}
  \label{phaseising}
\end{figure}

\begin{figure}
  \includegraphics[width=5cm]{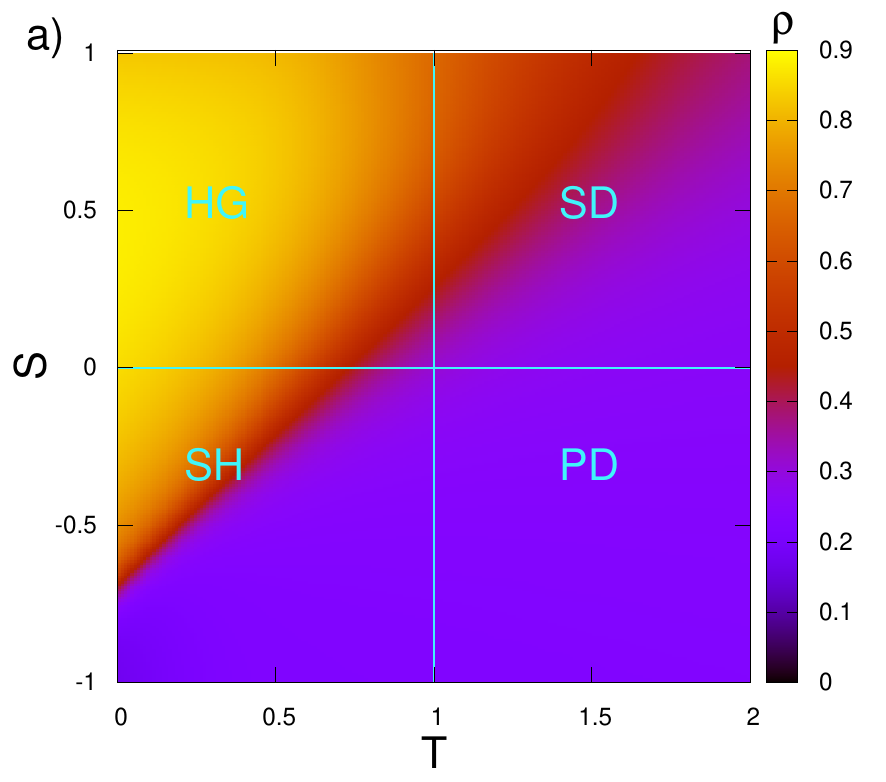}
    \includegraphics[width=5cm]{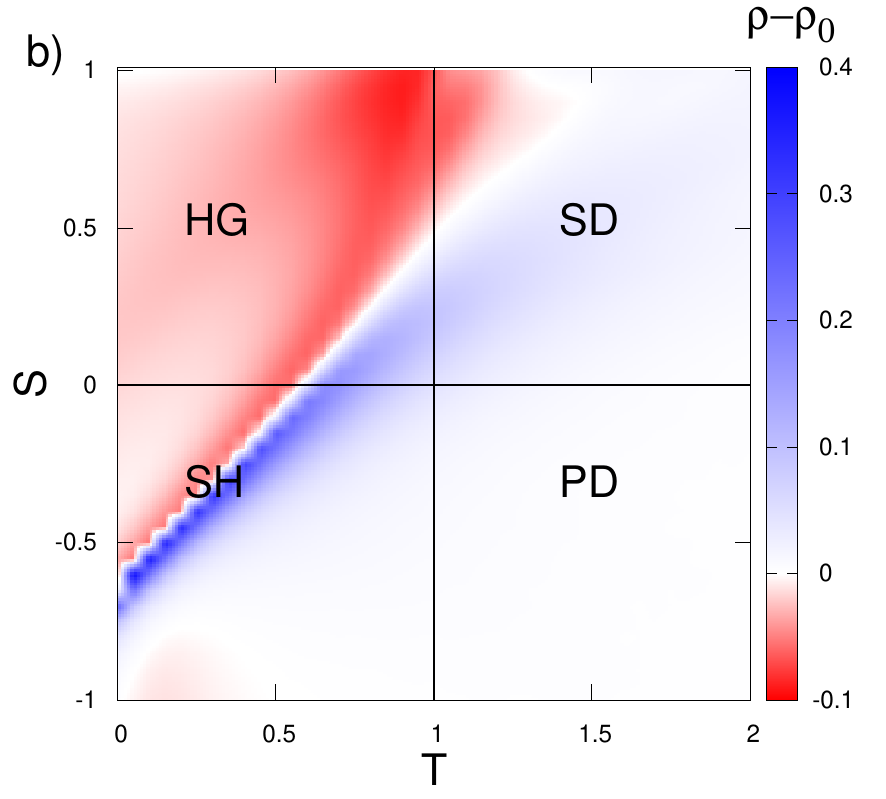}
  \caption{$T \times S$ parameter space for the WSLS model. Final cooperation fraction is represented in colours. a) high perturbation ($D=0.5$) and b) difference $(\rho-\rho_0)$.}
  \label{phasewsls}
\end{figure}

To conclude, we present the final average cooperation level for the whole parameter space $T \times S$ considering all update rules, i.e. Imitative (Figure~\ref{phaseimit}), Ising (Figure~\ref{phaseising}) and WSLS (Figure~\ref{phasewsls}). For all figures, we report the case for a high perturbation ($D=0.5$) in a) and the difference between the perturbed and unperturbed case, i.e. $(\rho-\rho_0)$,  in b) for the whole parameter space. We note that the parameter space for the unperturbed case can be found in the literature in~\cite{Amaral2016b} for the Win-Stay-Lose-Shift, in~\cite{Amaral2017} for the Ising model, and in~\cite{Szabo2016a} for the imitative model.

This analysis indicates that the perturbation changes the population dynamics mainly near phase transition regions. By using all three updating rules, we can observe the general effect that perturbation increases cooperation in the direction of more selfish games (i.e. high $T$, low $S$), while it does the opposite for altruistic games. We also note that the detrimental effect on cooperation is usually lower than the benefit for regions with high $T$ and a low $S$.
Eventually, we deem especially interesting that by using the Ising rule (see Figure~\ref{phaseising}), the perturbation increases the anti-coordination area linearly with the perturbation strength (top right region of the $T\times S$ parameter space).

\section{Conclusions}
\label{sec:Conclusions}
In this work, we studied payoff perturbations considering simple social dilemmas, with the main goal to clarify and quantify their effects when only the off-diagonal is interested, following the results reported in~\cite{amaral_javarone_2020}.
Beyond analysing perturbative methods in evolutionary games, the proposed model has the potential to actively contribute to a vivid debate, i.e. that on the relationship between heterogeneity and cooperation.
We emphasise that the form of heterogeneity we consider is related to that of risk, and reward, perception, i.e. an aspect covering a fundamental role in many social systems, and that might also be relevant, using a different interpretation, in other contexts.

Then, while~\cite{amaral_javarone_2020} focused on a general range of effects resulting from perturbed payoffs, here we concentrate on the strategy equilibrium reached by a population in presence of perturbations acting only on the rewards of betraying and the risks involved in getting betrayed (i.e. the temptation and the sucker's payoff). Hence, cooperators playing with cooperators, and defectors playing with defectors, are not affected by perturbations. In addition, here we analyse a wider range of game parameters, i.e. near and far phase transition points, in order to identify those values that allow perturbations to actively affect the dynamics of the game.

The evolution of strategies has been studied by Monte Carlo simulations, arranging players on a square lattice with periodic boundary conditions. To verify if the observed effects were robust in different settings, we considered three different update rules, Imitation, Win-Stay-Lose-Shift and Ising. We also analised the effects of Gaussian and uniform noise distributions.

Results show that perturbations can have more pronounced effects on the population dynamics mainly near the phase transitions. Notably, the average evolution of a population showed that the perturbations act very quickly, affecting mainly the initial evolution. Also, this effect is proportional to the perturbation amplitude when $T>1$. At the same time, the dampening in cooperation observed for $T<1$ seems to be very small in comparison.
In addition, the overall effect of the perturbation seems to be independent of the specific choice of the update rule, highlighting its robustness.
Then, the analysis related to the combined effect of heterogeneity and irrationality showed that these two perturbation sources are quite independent each other.

In order to understand the microscopic mechanism responsible for the cooperation enhancement near phase transitions, we studied the sub-population of cooperators and defectors that, locally, played more fraternal or egotistic games at each time-step. This is in line with previous works, which proposed that the benefit of diversity would be related to the random changes in the game class (Prisoner's Dilemma changing to Stag-hunt or Snow-Drift)~\cite {Perc2006, Perc2006a, Perc2006b}. This approach allowed to unveil what was responsible for the general effect observed in all cases. On average, we found that the payoff perturbation offers no unilateral contribution for more egotistic of fraternal games. Even so, it can locally promote cooperation for players that had a more positive perturbation. This effect can lead to the formation of stronger cooperative clusters, while defectors cannot benefit from said phenomena in the long run as they do not benefit from mutual support (see also~\cite{Javarone2016b}). On the other hand, in~\cite{tanimoto_pre07b} authors presented arguments on the irrelevance of game class changes for some contexts. While this seems to be an open topic, our initial results indicate that the random fluctuations have deep effects on the phase transitions of the model, being able to improve cooperation thanks to the unusual states found near phase transition points.
 
Summarising, our main finding is that off-diagonal perturbations cannot be considered as a trivial form of noise, as they seem to selectively enhance cooperation when defection is pervading the population, while they are able to support defection as the population begins to cooperate.
Therefore, in our view, the proposed model offers interesting insights on a relevant aspect of heterogeneity, not limited only to the studying of social phenomena. Notwithstanding, since several details might deserve further attention, we report those we consider potentially more interesting. For instance, the proposed model could be analysed considering populations arranged on complex networks~\cite{Wu2017, Battiston2017}, social behaviours like conformism~\cite{galam_ijmpc08, Galam2004, Javarone2016c} could be combined with random perturbations and, as suggested in~\cite{Stollmeier2018}, also some clinical investigations could find useful to represent specific biological phenomena by the proposed method. 
Finally, we see as especially interesting the connection of the payoff perturbation with recent works connecting evolutionary game theory with the Hamiltonian description of physical systems ~\cite{Nakamura2019, Szabo2016a}. These effects could be further studied in the light temporal Griffiths phases~\cite{Vazquez2011, Fiore2018}. Such an exotic state is known to appear near phase transitions in epidemiological and magnetic systems with directed percolation universality class, which is the same phase transition as the imitative model~\cite{Szabo2007}.

\begin{acknowledgments}
This research was supported by the Brazilian Research Agency CNPq (proc. 428653/2018-9).
\end{acknowledgments}

\end{document}